\documentclass[paper,notoc]{JHEP3}
\usepackage{cite}
\usepackage{graphicx}
\usepackage{amssymb,amsfonts,amsmath,bbold,latexsym}

\newcommand{\ep}{\varepsilon}

\preprint{ANL-HEP-PR-10-6}

\title{\boldmath 
Color-octet scalar effects on Higgs boson production in gluon fusion}

\author{Radja Boughezal\\
Institute for Theoretical Physics, University of Zurich,\\ Winterthurerstr. 190, 8057 Zurich, Switzerland \\ 
  E-mail: \email{radja@physik.uzh.ch}}
\author{Frank Petriello\\
 High Energy Physics Division, Argonne National Laboratory, Argonne, IL 60439, USA\\and\\
 Department of Physics, University of Wisconsin, Madison, WI 53706, USA\\
  E-mail: \email{frankjp@physics.wisc.edu}}

\abstract{ 
We compute the next-to-next-to-leading order QCD corrections to the gluon-fusion production of a Higgs boson in models with 
massive color-octet scalars in the ${\bf (8,1)_0}$ representation using an effective-theory approach.  We derive a compact analytic expression for the relevant
Wilson coefficient, and explain an interesting technical aspect of the calculation that requires inclusion of the quartic-scalar interactions at 
next-to-next-to-leading order.  We perform a renormalization-group 
analysis of the scalar couplings to derive the allowed regions of parameter space, and present phenomenological results for both the Tevatron and the LHC.  The modifications of the Higgs production cross section are 
large at both colliders, and can increase the Standard Model rate by more than a factor of two in allowed regions of parameter space.  We estimate that stringent constraints on the color-octet scalar parameters can be obtained using the Tevatron exclusion limit on Higgs production.

} 

\begin{document}

\section{Introduction}
\label{sec:introduction}

The Higgs boson is the last undiscovered particle of the Standard Model (SM).  The hunt for the Higgs boson to uncover its role in electroweak symmetry breaking is being undertaken at the Tevatron, and will soon begin at the Large Hadron Collider (LHC).  The CDF and D0 collaborations at the Tevatron have recently announced a 95\% exclusion limit on a SM Higgs boson with a mass in the range $162 \, \text{GeV} \leq m_h \leq 166 \, \text{GeV}$~\cite{Aaltonen:2010yv}, while the LEP collaborations have established the limit $m_h \geq 114$ GeV~\cite{Barate:2003sz}.  A crucial component of this search is the derivation of accurate theoretical predictions for the cross section for Higgs production.   The dominant hadronic production mechanism, gluon fusion through a top-quark loop, is known exactly through next-to-leading order in perturbative QCD~\cite{Djouadi:1991tka,Spira:1995rr}.   In the effective theory with $m_t \to \infty$, both the NLO corrections~\cite{Dawson:1990zj} and the NNLO corrections are known~\cite{Harlander:2002wh,Anastasiou:2002yz,Ravindran:2003um}.  When normalized to the full $m_t$-dependent leading-order result, this effective theory reproduces the exact NLO result to better than 1\% for $M_H < 2m_t$ and to 10\% or better for Higgs boson masses up to 1 TeV~\cite{ztalks}.  The QCD radiative corrections drastically alter the Higgs production cross section prediction; for example, the gluon fusion cross section is increased by roughly a factor of three above the LO prediction at the Tevatron after the NNLO corrections are included.  Only at NNLO is an accurate prediction free from debilitating uncertainties obtained.  Updated cross sections for Higgs production in gluon fusion for use at the Tevatron and LHC are available in Refs.~\cite{Anastasiou:2008tj,deFlorian:2009hc}.  For a recent review of the status of theoretical predictions for Higgs boson production in the SM, see Ref.~\cite{Boughezal:2009fw}.

The properties of the Higgs boson can be modified  in theories with additional particles, and measurement of these properties consequently serves as a window into physics beyond the SM.  In regions of supersymmetric parameter space, the production mechanism $gg \to h \to \gamma\gamma$ can be changed by up to a factor of two~\cite{Low:2009nj}.  In extra-dimensional theories, mixing between the Higgs and the scalar radion can modify the Higgs production rates by orders of magnitude~\cite{Hewett:2002nk}.  The gluon-fusion mode is particularly sensitive to new states, since they can contribute to the cross section at the same one-loop order as the SM particles.  It is therefore important to accurately predict the gluon-fusion production cross section in theories beyond the SM to assure that signatures of new physics are conclusively identified.

Several interesting extensions of the SM introduce new scalar states transforming in the adjoint representation under the color gauge group.  It was shown in Ref.~\cite{Manohar:2006ga} that a scalar particle transforming as $(\bf{8},\bf{2})_{1/2}$ under $SU(3) \times SU(2) \times U(1)$ is the only new scalar representation with non-trivial electroweak quantum numbers that can couple to quarks without introducing additional flavor violation beyond that present in the SM.  As the scalars can couple to the SM Higgs at tree-level, such particles can modify the gluon-fusion cross section through their propagation in loops.   The induced shifts to the $gg \to h$ cross section induced by $(\bf{8},\bf{2})_{1/2}$ scalars were studied at NLO in Ref.~\cite{Bonciani:2007ex}, and were found to be large.  Scalars in the $(\bf{8},\bf{1})_{0}$ representation can arise in theories with universal extra dimensions~\cite{Dobrescu:2007xf,Dobrescu:2007yp} and in technicolor models~\cite{Hill:2002ap}.  The primary decay for such states is expected to be into either $b\bar{b}$ or $t\bar{t}$, depending on the scalar mass and other model parameters.  The Tevatron experiments can search for these states via pair production of $(\bf{8},\bf{1})_{0}$ scalars, leading to a four $b$-jet final state.  The search reach in the scalar mass was recently estimated to be 280 GeV~\cite{Dobrescu:2007yp}.  The direct search for these scalars is rendered difficult by the large QCD background.  Reduction of the background requires significant cuts that reduce the signal and therefore the search reach.  It is possible that indirect searches for these scalars, such as via their influence on the Higgs production cross section, can probe masses competitive with direct searches.

We compute in this 
manuscript the NNLO corrections to the Higgs boson production cross section in models with a $(\bf{8},\bf{1})_{0}$ scalar.  To be as independent as possible from the origin of this scalar, we study a simple model that couples this scalar to both QCD and the Higgs doublet via renormalizable operators.  We utilize the effective-theory approach valid when both the SM top quark and the new scalar are heavier than the Higgs boson.  As explained above, we expect this to be an excellent approximation for Higgs masses in the interesting range.  We focus here on the $(\bf{8},\bf{1})_{0}$ state as a first case because it leads to a significant technical simplification when handling the effects of the scalar-sector potential.  As we explain later, the need to include the scalar self-couplings first arises at NNLO.  We compute the Wilson coefficient describing the interaction of the Higgs boson to gluons mediated by both the top quark and the color-octet scalar through NNLO in the QCD coupling constant.  We derive all renormalization constants required in the scalar sector.  All results are presented in compact analytic expressions.  We derive the renormalization-group equations governing the evolution of the scalar-sector couplings, and use them to determine the likely ranges of the various parameters which appear.  

In addition to the analytic results described above, we study the phenomenological implications of the color-octet scalar for the Higgs production cross section at both the Tevatron and the LHC.  Only at NNLO is the scale dependence sufficiently reduced to allow precise predictions for the scalar-induced effects to be obtained.  The deviations from the Standard Model prediction for Higgs production are large at both colliders.  Shifts in the Higgs cross section larger than the errors coming from scale variations and parton distribution functions are obtained for scalar masses approaching 1 TeV.  Deviations of a factor of two are obtained for scalar masses near the estimated direct search reach of $m_S \approx 300$ GeV.  We therefore believe that the current Tevatron exclusion limits on the Higgs boson would yield constraints on the color-octet parameter space competitive with direct-search constraints.  The indirect constraints depend on 
an undetermined Higgs-scalar coupling, and are therefore more model dependent.  However, from a low-energy perspective there is no symmetry reason to expect a small value for this coupling, and therefore a large region of allowed parameter space would be tested.

Our paper is organized as follows.  We describe our model for the $(\bf{8},\bf{1})_{0}$ scalar in Section~\ref{sec:model}.  In Section~\ref{calc} we describe our calculation, and present analytic results for the Wilson coefficient and all required renormalization constants in the scalar sector.  Numerical results for both the 
Tevatron and LHC are presented in Section~\ref{sec:num}.  We conclude in Section~\ref{sec:conc}.

\section{Details of the Model}
\label{sec:model}

We begin with the following Lagrangian, which describes the Standard Model coupled to a color-octet scalar in the $(\bf{8},\bf{1})_{0}$ representation:
\begin{eqnarray}
\label{eq:Lfull}
{\cal L}^{full} &=& {\cal L}_{SM}
         + \text{Tr}\left[D_{\mu} S D^{\mu} S\right]  
         - m_S^{'2} \,\text{Tr} \left[ S^2 \right]
         - g_s^2 \,G_{4S} \,\text{Tr} \left[ S^2 \right]^2
         -  \lambda_1 H^{\dagger}H \,\text{Tr} \left[ S^2 \right]
\nonumber \\
	&&-\lambda_h \left(H^{\dagger}H - \frac{v^2}{2}\right)^2.
\end{eqnarray}
$S$ denotes the matrix-valued scalar field $S = S^A T^A$, $H$ indicates the Higgs doublet before electroweak symmetry breaking, $v$ is the Higgs vacuum-expectation value, and $D_{\mu}$ is the covariant derivative for adjoint fields.  The Higgs quartic coupling has been explicitly included to define its normalization.  It will be used later when deriving the allowed range of the scalar couplings.  After electroweak symmetry breaking, the Higgs doublet is expanded as $H = \left(0,(v+h)/\sqrt{2} \right)$ in the unitary gauge.  The mass of the color-octet scalar becomes $m_S^2 = m_S^{'2} + \lambda_1 v^2/2$.  The Feynman rules which describe the scalar couplings to the Higgs boson $h$ and to gluons are easily obtained from Eq.~(\ref{eq:Lfull}).  The free parameters which govern the scalar properties are $m_S$, $\lambda_1$, and $G_{4S}$.  We note that higher-order operators which break the $S \to -S$ symmetry present above and which allow the scalar to decay are obtained in explicit models which contain this state~\cite{Dobrescu:2007xf,Dobrescu:2007yp}.  We neglect them here since we anticipate that they have little effect on the $gg \to h$ production cross section.

\begin{figure}[htbp]
   \centering
   \includegraphics[width=0.50\textwidth,angle=0]{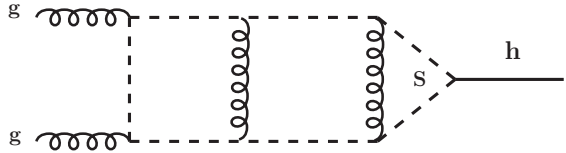}
   \caption{Example of a three-loop diagram necessitating the inclusion of the quartic-scalar term in the Lagrangian.}
   \label{scalar3lp}
\end{figure}

Before continuing, we comment on the appearance of the $\text{Tr} \left[ S^2 \right]^2$ term in the scalar potential.  A quartic-scalar coupling is generated by QCD interactions even if it is set to zero at tree-level.  At NNLO the quartic coupling must be included to obtain a renormalizable result due to the contributions of three-loop diagrams such as shown in Fig.~\ref{scalar3lp}.  We include this operator in the tree-level Lagrangian with a coefficient scaled by $g_s^2$, the QCD coupling constant squared, to permit an easier power-counting of loops.

\section{Calculational details and analytic results}
\label{calc}

\begin{figure}[htbp]
\centerline{
   \includegraphics[width=0.45\textwidth,angle=0]{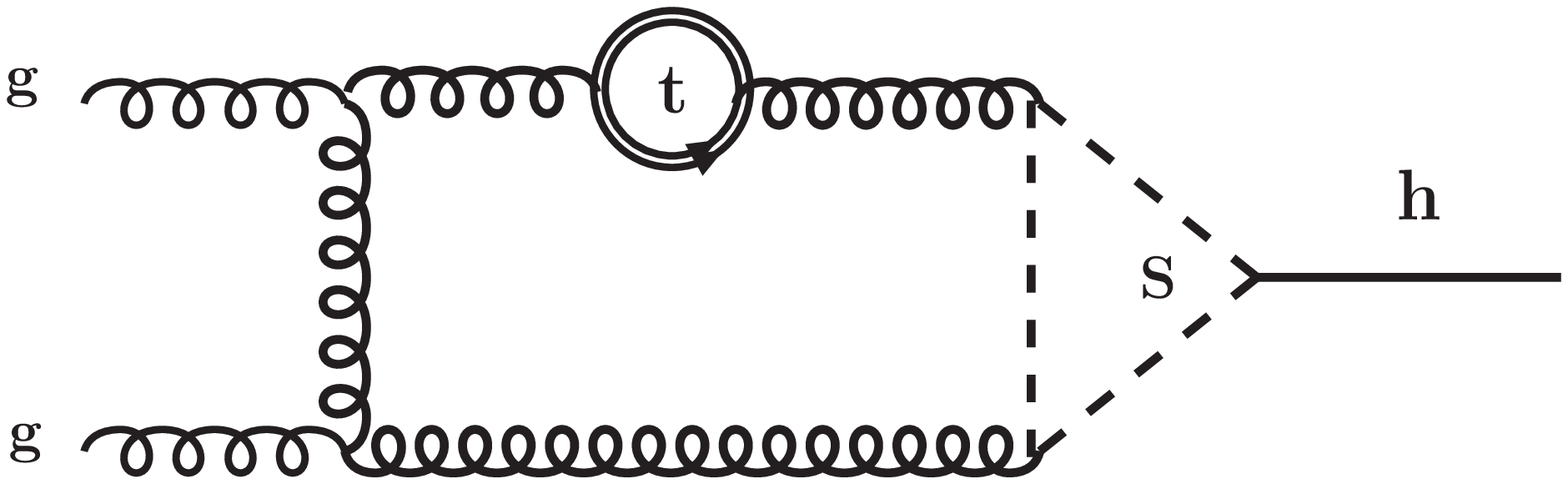}
    \includegraphics[width=0.45\textwidth,angle=0]{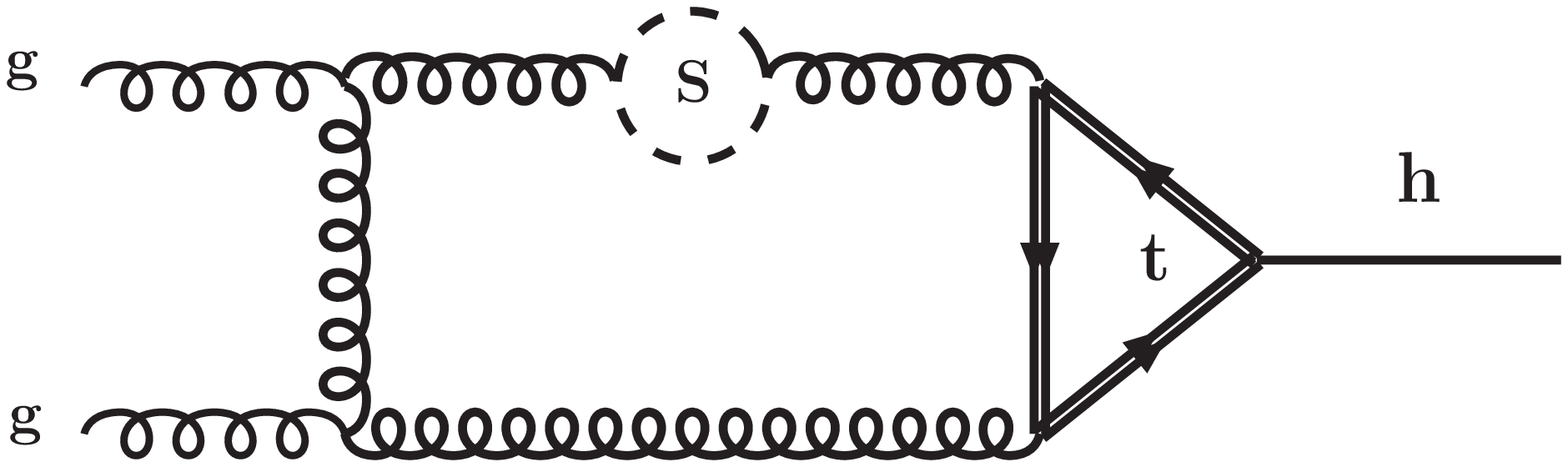}}
   \caption{Example two-scale diagrams contributing to the Wilson coefficient at NNLO.}
   \label{twoscale}
\end{figure}

We discuss here our derivation of the effective Lagrangian describing the interaction of the Higgs boson with gluons through NNLO in the QCD coupling constant.  When both the adjoint scalar and the top quark are heavier than the Higgs boson, they can be integrated out to derive the following effective Lagrangian:
\begin{equation}
\label{eq:Leff}
{\cal L}^{eff} = {\cal L}_{QCD}^{n_l,eff}
         - C_1 \,\frac{H}{v} \,{\cal O}_1,
\end{equation}
where $C_1$ is a Wilson coefficient and the operator ${\cal O}_1$ is
\begin{equation}
{\cal O}_1 = \frac{1}{4} \,{G'}^a_{\mu\nu} {G'}^{a\mu\nu} \, .
\end{equation}
The effective Lagrangian for the gluons and light quarks, ${\cal L}_{QCD}^{n_l,eff}$, has the same form as ${\cal L}_{QCD}^{n_l}$ except that the fields and parameters it contains are rescaled by decoupling constants that account for the effects of the heavy states. In order to distinguish the fields and parameters occurring in the full Lagrangian from those in the effective Lagrangian, we denote the latter ones with a prime.  In this manuscript we integrate out the top quark and the color-octet scalar in a  single step.  Our calculation is therefore a two-scale problem.  This is clearly demonstrated by the example diagrams contributing to the NNLO Wilson coefficient shown in Fig.~\ref{twoscale}.  If a large hierarchy exists between the scalar and top masses, a two-step procedure in which the scalar and top quark are integrated out separately can instead be used.

The Wilson coefficient $C_1$ can be constructed by computing the amplitude for $gg \to h$ in the limit that the initial gluon momenta vanish, as reviewed in Ref.~\cite{Steinhauser:2002rq}.  We start from the relation
\begin{equation}
\label{eq:C01}
\frac{\zeta^0_3 C^0_1}{v} =
\frac{\delta^{a_1 a_2}
 \left( g^{\mu_1\mu_2} (p_1\cdot p_2) - p_1^{\mu_2} p_2^{\mu_1} \right)}
     { (N^2-1) (d-2) {(p_1\cdot p_2)^2} }
     {\cal M}^{0,a_1 a_2}_{\mu_1\mu_2}(p_1, p_2)
     \bigr\rvert_{p_1 = p_2 = 0}
\end{equation}
between the bare Wilson coefficient $C^0_1$ and the bare amplitude for the process $g g \to H$ in the full theory. Here, $p_1$ and $p_2$ are the momenta of the two gluons, $N$ is the number of colours, and $d=4-2\ep$ is the dimension of space-time. The factor $\zeta^0_3$ is the bare decoupling coefficient by which the bare gluon fields in the full and effective theories are related:
\begin{equation}
{G'}^{0,a}_{\mu} = \sqrt{\zeta^0_3} \,{G}^{0,a}_{\mu} \, .
\end{equation}
It can be expressed as
\begin{equation}
\label{eq:zeta03Pi}
\zeta^0_3 = 1 + \Pi^0_G(p=0) \, ,
\end{equation}
where $\Pi^0_G$ is the transverse part of the bare gluon self-energy in the full theory. In order to obtain the Wilson coefficient through NNLO in the QCD coupling constant, ${\cal M}^{0,a_1 a_2}_{\mu_1\mu_2}$ is needed up to three loops, while $\Pi^0_G(p=0)$ is needed through two loops.  We can set all scaleless integrals to zero in dimensional regularization, and as a result, only diagrams containing at least one massive scalar propagator contribute to $\Pi^0_G(p=0)$ and to the right hand side of Eq.~(\ref{eq:C01}).  We perform our calculations in a covariant gauge with gauge parameter $\xi$, leading to the following gluon propagator:
\begin{equation}
\frac{i}{q^2} \left( -g^{\mu\nu} + \xi \frac{q^{\mu}q^{\nu}}{q^2} \right) \, .
\end{equation}
In all diagrams, terms up to first order in $\xi$ are retained.  All $\xi$-dependent terms cancel in $\Pi^0_G(p=0)$ up to two-loop order, and in ${\cal M}^{0,a_1 a_2}_{\mu_1\mu_2}$ up to three-loop order, demonstrating that $C^0_1$ is gauge independent.

Initially, all quantities appearing in eq.~(\ref{eq:C01}) are expressed in terms of the bare masses $m_S^0$, $m_T^0$, and the bare coupling constants $g_s^0$, $G_{4S}^0$.  We derive here the renormalization constants that relate the bare parameters to physical ones.  The scalar and top-quark masses are renormalized in the $\overline{MS}$ scheme according to
\begin{equation}
\label{eq:mSren}
m_S^0 = \sqrt{Z_{m_S}}\, m_S,  \;\;m_T^0 = Z_{m_T}\, m_T,
\end{equation}
with
\begin{equation}
\label{eq:Zms}
\begin{split}
Z_{m_S} &= 1 - \frac{a}{4\,\ep}\,(9 - 10\,G_{4S}) 
 \\&
   + \,a^2 \,\left( \frac{237 - 6 \,n_l- 360\, G_{4S} 
   + 260\, G_{4S}^2 }{32\, \ep^2} 
   + \frac{-53 + 10 \,n_l+ 480\, G_{4S} - 100\, G_{4S}^2 
   - 72 \,\,m_T^2/m_S^2}{64 \,\ep}\right)\,,
\end{split}
\end{equation}
and
\begin{equation}
\label{eq:ZmT}
Z_{m_T} = 1 - \frac{a}{\ep} 
 \,+ \,a^2 \,\left( \frac{83 - 4\, n_l}{48\, \ep^2} 
   + \frac{-553 + 20\,n_l}{288 \,\ep}  \right)\,.
\end{equation}
We note that the appearance of the $m_T^2/m_S^2$ term in the expression for $Z_{m_S}$ is not unexpected.  The scalar mass receives an additive contribution from the top quark since no symmetry prevents it, and as a result of attempting to force a multiplicative renormalization in Eq.~(\ref{eq:mSren}) this mass ratio appears in $Z_{m_S}$.  Equivalently, one can formulate the renormalization by introducing a matrix of constants: 
\begin{equation}
(m_i^0)^2 = \sum_{j=T,S} Z_{ij} (m_j^0)^2,
\end{equation}
with $i=T,S$.  The appearance of the $m_T^2 /m_S^2$ term in Eq.~(\ref{eq:mSren}) is then converted into the presence of an off-diagonal term $Z_{ST}$ in this matrix. 

The quartic coupling first appears in the $gg \to H$ amplitude at the two-loop level, and its renormalization is therefore only required to one loop:
\begin{equation}
\label{eq:G4Sren}
G^0_{4S} = Z_{4S}\, G_{4S}\,,
\end{equation}
with
\begin{equation}
\label{eq:Z4S}
Z_{4S} = 1 + \frac{a}{\ep}\,\left(
        -\frac{49}{24} + \frac{27}{16\,G_{4S}} + 4\,G_{4S}
      - \frac{n_l}{6}\right)\,.
\end{equation}
We convert the bare strong coupling constant of the full theory into the bare coupling of the effective theory using decoupling constants obtained from the ghost self energy and the ghost-gluon vertex up to two-loops, as described in detail in~\cite{Chetyrkin:1997un,Steinhauser:2002rq}.  We then renormalize it in the effective theory using the $\overline{MS}$ scheme:
\begin{equation}
g'^0_s = \mu^{\ep}\, Z'_g\, g'_s \, ,
\end{equation}
where
\begin{equation}
Z'_g = 1 - a'\, \frac{\beta_0}{2\,\ep}
         + a'^2 \,\left( \frac{3\,\beta_0^2}{8\,\ep^2}
                     -\frac{\beta_1}{4\,\ep} \right)\,.
\end{equation}
Here,
\begin{equation}
a' = \frac{\alpha_s^{(n_l)}(\mu)}{\pi} = \frac{{g'_s}^2}{4\,\pi^2}
\end{equation}
and the first two coefficients of the $\beta$ function are given by
\begin{eqnarray}
\beta_0 & = & \frac{1}{4} \,\left( 11 - \frac{2}{3} \,n_l \right)\,,
\\
\beta_1 & = & \frac{1}{16} \,\left( 102 - \frac{38}{3}\, n_l \right)\,.
\end{eqnarray}
The Wilson coefficient itself requires a renormalization factor~\cite{Spiridonov:1984br,Spiridonov:1988md,Chetyrkin:1996ke}:
\begin{equation}
C_1 = \frac{1}{Z_{11}} \,C^0_1\,,
\end{equation}
with
\begin{equation}
\frac{1}{Z_{11}} = 1 + a' \,\frac{\beta_0}{\ep} + a'^2 \,\frac{\beta_1}{\ep}\,.
\end{equation}
We split our final result for the renormalized Wilson coefficient into three contributions:
\begin{equation}
C_1 = C_{TTH}+C_{SSH}+C_{TS}.
\end{equation}
The known SM Wilson coefficient $C_{TTH}$ arising from just the top-quark contributions is given by~\cite{Chetyrkin:1997iv,Chetyrkin:1997un,Kramer:1996iq}
\begin{equation}
\label{eq:CTTH}
C_{TTH} = -\frac{a'}{3} - \frac{11 \,a'^2}{12} + 
 a'^3 \,\biggl[\frac{1}{864}\, \left(-2777 + 684 \,L_T\right) + 
        \frac{1}{288}\, \left(67 + 64 \, L_T\right)\, n_l\biggr]\,.
\end{equation}
The contribution $C_{SSH}$ arises from diagrams involving the coupling of the adjoint scalars to the Higgs boson.  Examples of such contributions are given by Fig.~\ref{scalar3lp} and the left-most diagram in Fig.~\ref{twoscale}.  The full contribution to $C_{SSH}$ is given by
\begin{eqnarray}
\label{eq:CSSH}
C_{SSH} &=& -\frac{\lambda_1 v^2}{2 \,m_S^2}\,
           \Biggl\{
                 \frac{a'}{4} 
                 + a'^2\,\biggl[\frac{33}{16} + \frac{5\,G_{4S}}{8}\biggr]
                 + a'^3\,\biggl[ n_l\,\left(\frac{-101}{288} +
                   \frac{7\,L_S}{24}\right) 
\\ \nonumber
                &+& G_{4S}^2\,\left(\frac{-35}{16} + 5\,L_S \right) 
                 + \frac{9\,L_S\,(-43 +8\,x^2)}{64}
                 - \frac{3\,(76 - 3895\,x^2 + 257\,x^4)}{1024\,x^2} 
\\ \nonumber
                &-& G_{4S}\,\left(\frac{-705}{64} + \frac{575\,L_S}{96} 
                 + \frac{5\,\mbox{ln}(x)}{24}\right) 
    + \frac{3\,\left(76 + 37\,x^2 + 86\,x^4 + 225\,x^6\right)\,}{2048\,x^3}\;
\times
\\ \nonumber &&
\\ \nonumber &&  \qquad\;\;
       \biggl(\mbox{Li}_3(x) - \mbox{Li}_3(-x) \biggr)
\\ \nonumber &&
\\ \nonumber
                &+& \mbox{ln}^2(x)\,
                \biggl\{-\frac{-228 + 41\,x^2 - 192\,x^4 + 675\,x^6}
                      {2048\,(-1 + x)\,x^2\,(1 + x)} 
    + \frac{3\,\left(76 + 37\,x^2 + 86\,x^4 + 225\,x^6\right)}{4096\,x^3}\,
\times
\\ \nonumber &&
\\ \nonumber &&  \qquad\quad
     \biggl(\mbox{ln}(1 +x) - \mbox{ln}(1 -x)\biggr) \biggr\} 
\\ \nonumber &&
\\ \nonumber
    &+& 3\,\mbox{ln}(x)\,\biggl\{\frac{76 - 111\,x^2 + 159\,x^4}
       {1024\,x^2}
     - \frac{76 + 37\,x^2 + 86\,x^4 + 225\,x^6}{2048\,x^3}\,
      \biggl(\mbox{Li}_2(x) - \mbox{Li}_2(-x)  \biggr) \biggr\}
                 \biggr]
           \Biggr\}.
\end{eqnarray} 
The remaining contributions $C_{TS}$ come from the adjoint scalar propagating
in loops, but where the Higgs couples to the top quark.  This piece gives the
correction to the Higgs production cross section that would result if the
scalar-Higgs coupling were set to zero.  An example contribution is given by
the right-most diagram in Fig.~\ref{twoscale}.  
We used Ref.~\cite{Bekavac:2009gz} for the non-trivial 3-loop master integrals 
with two scales and four propagators. The full contribution begins first 
at the three-loop order, and is given by the following expression:
\begin{eqnarray}
\label{eq:CTS}
C_{TS} &=&  a'^3\,\biggl[ 
            \frac{9\,L_S\,x^2}{8} 
          - \frac{2052 + 1075\,x^2 + 1755\,x^4}{9216\,x^2} 
\\ \nonumber  
          &+& \mbox{ln}(x)\,\Biggl\{
            \frac{684 + 409 \,x^2 + 1431 \,x^4}{3072 \,x^2} 
          - \frac{3\, \left(76 + 37 \,x^2 + 86 \,x^4 + 225 \,x^6\right)
          }{2048\,x^3}\;\biggl(\mbox{Li}_2(x) - \mbox{Li}_2(-x) \biggr)\Biggr\}
\nonumber\\ \nonumber  
          &+& \mbox{ln}^2(x)\,\Biggl\{
           - \frac{-228 + 41 \,x^2 - 192\, x^4 + 675\, x^6}
                  {2048 \,(-1 + x)\, x^2 \,(1 + x)} 
           + \frac{3\, \left(76 + 37 \,x^2 + 86 \,x^4 + 225 \,x^6\right)
          }{4096\,x^3} \; \times
\\ \nonumber && \qquad\quad
            \biggl(\mbox{ln}(1 +x) - \mbox{ln}(1-x)\biggr)\Biggr\}
          + \frac{3\,\left(76 + 37\,x^2 + 86\,x^4 + 225\,x^6\right)\,}
           {2048\,x^3}\;\biggl(\mbox{Li}_3(x) - \mbox{Li}_3(-x) \biggr)
      \biggr]
\end{eqnarray} 
In these expressions, we have set $L_{i} = \mbox{ln}\left(m_{i}/\mu\right)$ for $i = (S,T)$,
$ x = m_T/m_S$, and we have used  $m_T$ and $m_S$ to denote the $\overline{MS}$ scalar
and top quark masses.

\section{Numerical Results}
\label{sec:num}

We now present numerical results for the Higgs boson production cross section in gluon fusion, to study the deviations induced by the color-octet scalar.  We include the effects of the top quark, the scalar, and also the bottom quark on the Higgs cross section.  We comment first on precisely what terms we include in the cross section. The leading-order amplitude for the $gg \to h$ process takes the form
\begin{equation}
\mathcal{A}^{LO} = \mathcal{A}^{LO}_T+ \mathcal{A}^{LO}_B + \mathcal{A}^{LO}_S,
\end{equation}
where the subscripts $T,B,S$ respectively denote the top, bottom, and scalar contributions. Upon squaring this amplitude, interferences between the contributions of each particle are obtained.  We denote by $\sigma_{T+S}$ the terms obtained by squaring together the top and scalar amplitudes, and keeping both the interference term and the pieces from each separate particle squared.  We let $\sigma_{TB},\sigma_{SB}$ denote the interferences between the bottom-quark amplitude with the top and the scalar pieces, respectively.  For the cross section at the $n$-th order in perturbation theory, we 
use the following expression:
\begin{equation}
\sigma^n = \sigma_{T+S}^{LO}(m_T,m_S) \, K^n_{EFT}+\sigma_{SB}^{LO}(m_S,m_b)+\sigma_{TB}^{LO}(m_T,m_b)+\sigma_{BB}^{LO}(m_b).
\end{equation}
$K^n_{EFT}$ denotes the ratio of the $n$-th order cross section over the LO result, with both computed in the effective theory defined in Eq.~(\ref{eq:Leff}).  The cross section multiplying the $K$-factor is the LO cross section with the exact dependence on the scalar and top-quark masses.  The remaining terms account for the scalar-bottom interference, the top-bottom interference, and the bottom-squared contribution at LO with their exact mass dependences.  In the SM, the scaling of the exact LO cross section by the EFT $K$-factor is known to furnish an approximation accurate to the few-percent level or better for Higgs masses below roughly 400 GeV~\cite{ztalks}.  For the scalar, this approximation has been studied against the exact NLO calculation~\cite{Bonciani:2007ex}, and is again accurate to the $1-2$\% percent level for $m_h \leq m_S$, which is the region we focus on here.  If desired, the exact NLO corrections to the top-bottom interference and bottom-squared 
terms~\cite{Spira:1995rr,Anastasiou:2009kn} can be included, as can those to the bottom-scalar interference~\cite{Bonciani:2007ex}.  These affect the cross section at the $1-2\%$ level, and for simplicity are neglected.  Various electroweak corrections known for the SM contribution~\cite{Aglietti:2004nj,Actis:2008ug,Anastasiou:2008tj,Keung:2009bs} are not known for the scalar, and for consistency are neglected.

We use the the pole mass $m_T = 173.1$ GeV for the top quark~\cite{:2009ec}, while for the bottom quark we use the $\overline{MS}$ mass with $\overline{m}_b (10 \, \text{GeV}) = 3.609$ GeV~\cite{Kuhn:2007vp}.  The choice of pole or $\overline{MS}$ mass for the $b$-quark has an insignificant effect on the
fractional deviation between the scalar-induced cross section and the
SM result.  We use the MSTW parton distribution functions~\cite{Martin:2009iq} extracted to the appropriate order in perturbation theory.  For the scalar sector, we must set the parameters $\lambda_1$, $G_{4S}$, and $m_S$.  We perform a renormalization-group analysis to constrain the possible values of the scalar couplings.  We describe this analysis below.

\subsection{Renormalization-group analysis of the scalar couplings}

While the scalar mass $m_S$ can be constrained by direct searches performed at the Tevatron~\cite{Dobrescu:2007yp}, such analyses do not restrict the allowed regions of the scalar couplings $G_{4S}$ and $\lambda_1$. To determine the reasonable range for these couplings, we instead derive constraints arising from 
theoretical consistency of the model.  For the potential in Eq.~(\ref{eq:Lfull}) to be bounded at large values of $\text{Tr}[S^2]$, we must have $G_{4S}>0$.  The coupling $\lambda_1$ may be either 
positive or negative.  For negative values, the restriction of a bounded potential at large field values imposes the constraint $|\lambda_1|<2 \sqrt{g_s^2 G_{4S} \lambda_h}$.  This restriction does not exist for $\lambda_1 >0$.  For negative values of $\lambda_1$ the cross section for Higgs production is decreased due to destructive interference between the top-quark Wilson coefficient and 
the scalar contribution in Eq.~(\ref{eq:CSSH}).  Since the allowed parameter space for positive $\lambda_1$ is larger, and also since Tevatron restrictions on positive $\lambda_1$ are stronger, 
we focus in this paper on this region.

We derive further constraints by demanding that the couplings do not encounter a Landau pole for energy scales up to some cutoff $\Lambda$.  At that scale, the model must be embedded into a more complete theory that removes the Landau pole.  Since the couplings increase in the ultraviolet, this restricts their values at lower energies relevant for studies at the Tevatron and the LHC.  As long as $\Lambda$ is smaller than the energies probed by experiment, the model defined in Eq.~(\ref{eq:Lfull}) provides a consistent framework for the scalar interactions.

We begin by presenting the system of coupled differential equations which governs the one-loop evolution of the $\overline{MS}$ couplings 
$G_{4S}$, $\lambda_1$, and $\lambda_h$.  We derive these results from the
one-particle irreducible Green's functions for the four-scalar interaction,
the interaction of two scalars with two Higgs bosons and the four-Higgs coupling.  The renormalization group equations are as follows:
\begin{eqnarray}
\label{eq:RGeq}
\frac{d G_{4S}}{dL} &=& 4 \,a\, G_{4S}^2 -\frac{49}{24} a\, G_{4S} -\frac{n_l}{6} a\, G_{4S} +\frac{27}{16} a + \frac{1}{64\pi^4} \frac{\lambda_1^2}{a}, \nonumber \\
\frac{d \lambda_1}{dL} &=& \frac{1}{8\pi^2} \lambda_1^2 - \frac{9}{4} a\, \lambda_1 + \frac{5}{2} a \,\lambda_1 G_{4S} + \frac{3}{8\pi^2} \lambda_1 \lambda_h , \nonumber \\
\frac{d \lambda_h}{dL} &=& \frac{1}{8\pi^2} \lambda_1^2 + \frac{3}{4\pi^2} \lambda_h^2,
\end{eqnarray}
where $L = \text{ln}\,\mu^2$, $a=g_s^2/(4\pi^2)$, and $n_l=5$ is the number of light fermions.  The running of the strong coupling $a$ including the scalar is well-known.  These equations can be checked to agree with well-known results available in the literature~\cite{Callaway:1988ya}.  For this part of our analysis only, we have permitted the Higgs boson to propagate in loops in order to derive the correct dependence of these equations on all scalar couplings.

\begin{figure}[htbp]
   \centering
   \includegraphics[width=0.50\textwidth,angle=90]{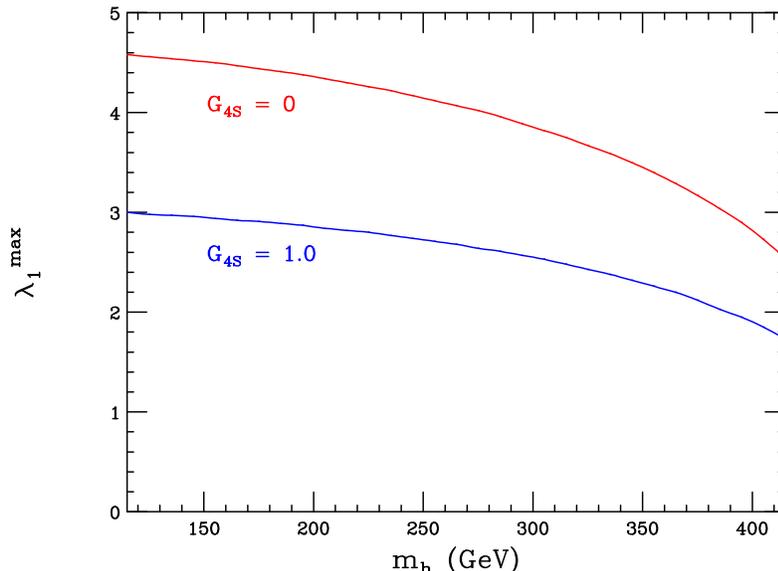}
    \vspace{-0.5cm}
   \caption{Maximum value of $\lambda_1$ allowed by perturbativity for the choices $G_{4S}(v)=1$ and $G_{4S}(v)=0$, as a function of Higgs mass.}
   \label{lam1max}
\end{figure}

We solve these equations numerically, starting the evolution at $\mu = v$ where the Higgs quartic coupling $\lambda_h$ is related to the Higgs boson mass via
$m_h^2 = 2 \lambda_h(v) v^2$.  We vary the starting values of the other couplings $\lambda_1(v)$ and $G_{4S}(v)$, and demand that all couplings remain perturbative until the 
scale $\Lambda = 10$ TeV.  Stronger restrictions would result if perturbativity were imposed up to a higher energy such as the grand-unified scale or Planck scale, but we use the weaker constraint here to assure all potentially relevant parameter space is included.  We show the results of our analysis below in Table~\ref{tab:landau} for the value $m_h=165$ GeV, where we fix one coupling to an allowed value and derive the restriction on the other.  In Fig.~\ref{lam1max} we show the maximum allowed value of $\lambda_1(v)$ obtained by demanding $\lambda_1(\mu)$ remain perturbative until $\mu=10$ TeV, as a function of the Higgs mass, for the choices $G_{4S}(v)=1$ and $G_{4S}(v)=0$.

\begin{table}[htbp]
  \begin{center}
    \begin{tabular}{|c|c||c|c|}
    \hline
    \multicolumn{2}{|c||}{Fixed $\lambda_1(v)$} &  \multicolumn{2}{c|}{Fixed $G_{4S}(v)$} \\
      \hline \hline
		$\lambda_1(v)=0$ & $G_{4S}(v) \leq 1.5$ & $\lambda_1(v) \leq 4.3$ & $G_{4S}(v) = 0$ \\ \hline
		$\lambda_1(v)=2.0$ & $G_{4S}(v) \leq 1.3$ & $\lambda_1(v) \leq 3.5$ & $G_{4S}(v) = 0.6$ \\ \hline
		$\lambda_1(v)=4.0$ & $G_{4S}(v) \leq 0.3$ & $\lambda_1(v) \leq 2.2$ & $G_{4S}(v) = 1.2$ \\ \hline
		 \hline
          \end{tabular}
  \end{center}
  \caption{Restrictions on the couplings $\lambda_1$ and $G_{4S}$ at the input scale $\mu=v$ arising from demanding perturbativity until $\mu=10$ TeV.  In the three left-most columns, $\lambda_1(v)$ is fixed to a given value, and the corresponding constraint on $G_{4S}(v)$ is derived.  In the three right-most columns, $G_{4S}(v)$ is fixed to a given value and a restriction on $\lambda_1(v)$ is found.  The value $m_h=165$ GeV is assumed.
      \label{tab:landau} }
\end{table}

We note that the combination $\lambda_1 v^2 /m_S^2$ is scale-invariant if only QCD-induced $\alpha_S$ corrections are considered (i.e., the Higgs boson is not allowed to propagate in loops).  This is a consequence of the low-energy theorem for Higgs production~\cite{Kniehl:1995tn}.  We have verified this by explicit calculation in the $\overline{MS}$ scheme.  If instead an on-shell scheme is chosen for the scalar mass, one should be able to implement a similar choice for $\lambda_1$ that preserves the renormalization-group invariance of the ratio. If that is done, the Wilson coefficient presented in Section~\ref{calc} takes the same form in both the $\overline{MS}$ and pole schemes, as the scalar mass appearing in $L_S$ and the top mass in $L_T$ can be interpreted as either the $\overline{MS}$ or pole masses.  The difference occurs only at NNNLO in perturbation theory, which is beyond the scope of our calculation.  In presenting our numerical results we interpret the masses of the top and scalar as pole masses, which assumes the choice of 
scheme for $\lambda_1$ described above.

In our numerical results for the Tevatron, we set $\lambda_1$ to the fixed value $\lambda_1 = 2.5$.  For LHC numbers we use $\lambda_1 = 1.5$ to allow results for larger Higgs masses consistent with perturbativity to be displayed.   We assume that the same scheme is chosen for both this quantity and the scalar mass so that the ratio is scale invariant.  We have checked that the contributions from $C_{TS}$ in Eq.~(\ref{eq:CTS}) are negligibly small, and the effects of the explicit mass dependence appearing in $L_S$ are at the percent level.  The cross section therefore depends primarily on the scalar parameters through the ratio $\lambda_1/m_S^2$.  Results for other values $\lambda_1^{o}$, but keeping the mass fixed at $m_S$, can be approximately obtained by studying the presented results at a mass $m_S^{o}$ given by $\lambda_1/m_S^2 = \lambda_1^{o} /(m_S^{o})^2$.  We also set $G_{4S}(v)=1.0$, and account for the evolution using the leading-order evolution equation in Eq.~(\ref{eq:RGeq}).  We neglect the dependence of $G_{4S}(\mu)$ on $\lambda_1$ for consistency with our calculation of the Wilson coefficient, where only QCD corrections were included and the Higgs was not allowed to propagate in loops.  Inclusion of this term has only a small effect on the running of $G_{4S}$.  The choice of $G_{4S}$ only affects the cross section at the few-percent level.  The coupling $\lambda_1$ always appears in the ratio $\lambda_1 v^2 /(m_S^2)$, as is clear from the presentation of the Wilson coefficient contribution $C_{SSH}$ in Eq.~(\ref{eq:CSSH}).  Varying the scalar mass therefore adequately accounts for possible cross-section deviations obtainable in this model.  

\subsection{Results for the Tevatron and the LHC}

\begin{figure}[ht]
   \centering
   \includegraphics[width=0.75\textwidth,angle=90]{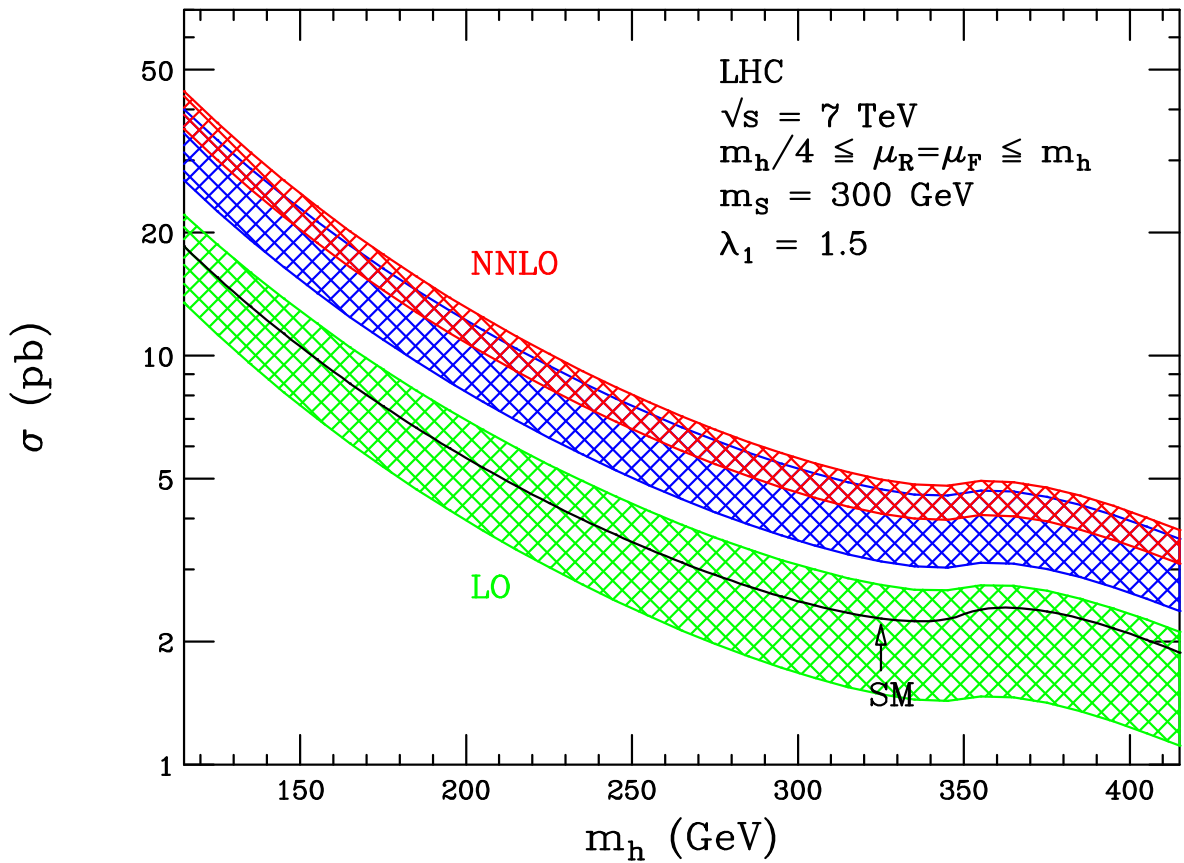}
   \vspace{-0.5cm}
   \caption{Higgs production cross section in gluon-fusion at the LHC for $m_S=300$ GeV as a function of the Higgs boson mass.  The bands indicate the scale variation $m_h/4 \leq \mu \leq m_h$.  From bottom to top, the bands indicate the variations of the LO, NLO, and NNLO cross sections.  All other parameters are as described in the text.  For orientation, the SM result at NNLO for the central value $\mu=m_h /2$ is shown.}
   \label{fig:mhplot}
\end{figure}

We now present numerical results for both the Tevatron and the LHC, for which we assume $\sqrt{s}=7$ TeV.  We begin by showing in Fig.~\ref{fig:mhplot} the LHC cross section for $m_S=300$ GeV at LO, NLO and NNLO in QCD perturbation theory, to see the effect of including higher-order QCD corrections.  The renormalization and factorization scales are equated to $\mu_F = \mu_R = \mu$, and are varied in the range $m_h/4 \leq \mu \leq m_h$, consistent with previous studies of the Higgs production cross section in the SM.  The scale-variation errors are large at both LO and NLO, and the corresponding error bands do not overlap.  Only at NNLO can a reliable prediction for the cross section be made.  It is also clear from this figure that large variations from the SM prediction are possible for scalar masses near the expected Tevatron limit of $m_S \approx 300$ GeV.  The cross section differs from the SM result by more than a factor of two for this parameter value.

To study further the effect of the color-octet scalar on the Higgs cross-section prediction at both the Tevatron and the LHC, we show below in Fig.~\ref{fig:TEVmsplot} and Fig.~\ref{fig:LHCmsplot} the cross sections for the example Higgs mass $m_h =165$ GeV as functions of the scalar mass at both the Tevatron and LHC.  Deviations from the SM are visible over scale errors at both colliders for scalar masses approaching 1 TeV.   The scalar contributions to the Higgs production cross section are large, and direct searches are hindered by the need to pair produce the scalars and by the large QCD background.  This suggests that the indirect constraints on the scalar parameter space coming from the Tevatron Higgs exclusion limit could be as strong as the direct search reach.

\begin{figure}[ht]
   \centering
   \includegraphics[width=0.75\textwidth,angle=90]{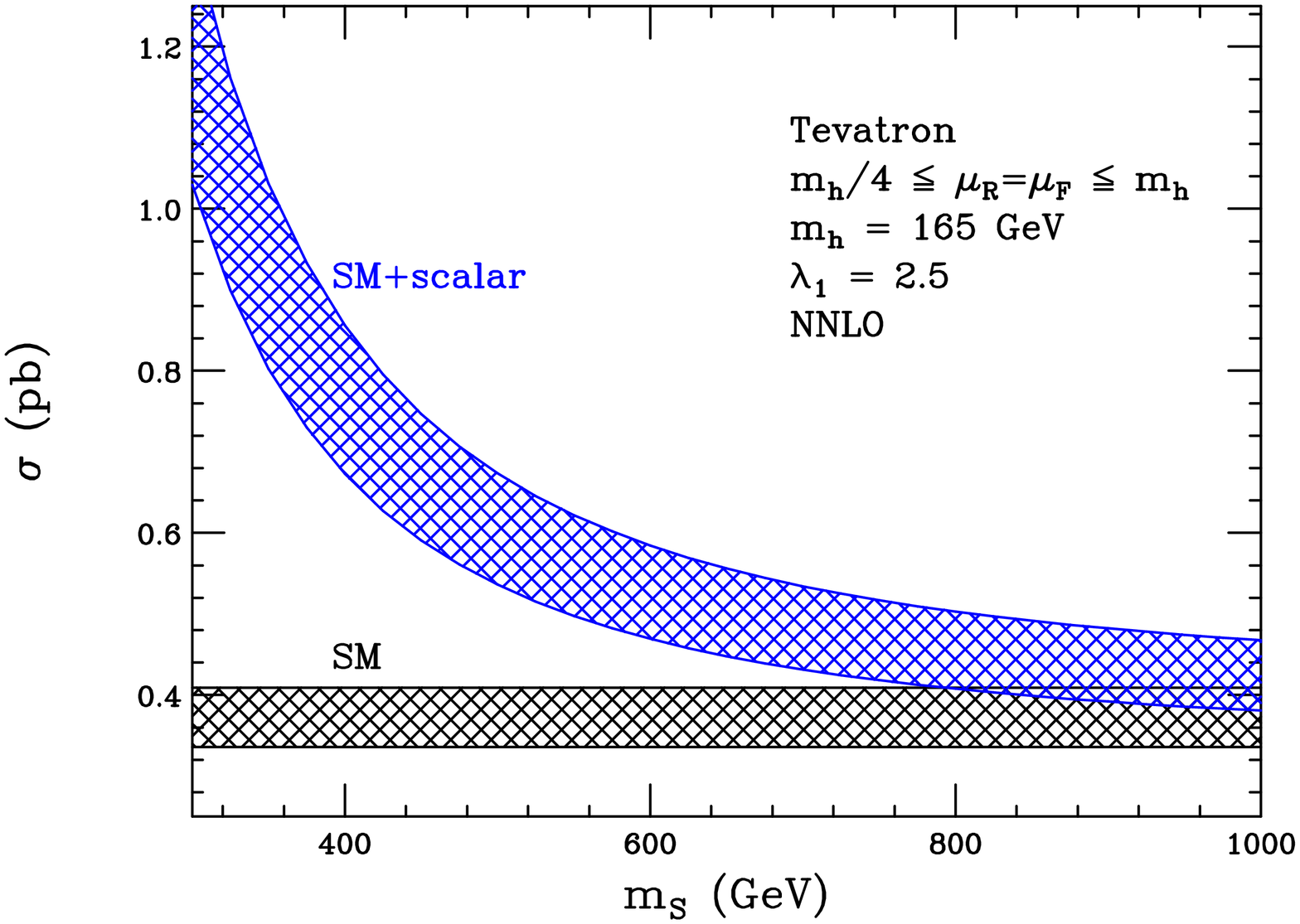}
   \vspace{-0.5cm}
   \caption{Higgs production cross section in gluon-fusion at the Tevatron for $m_h=165$ GeV as a function of the scalar mass.  The bands indicate the scale variation $m_h/4 \leq \mu \leq m_h$.  All other parameters are as described in the text.  Also shown is the SM cross section with its corresponding scale uncertainty.}
   \label{fig:TEVmsplot}
\end{figure}

\begin{figure}[ht]
   \centering
   \includegraphics[width=0.75\textwidth,angle=90]{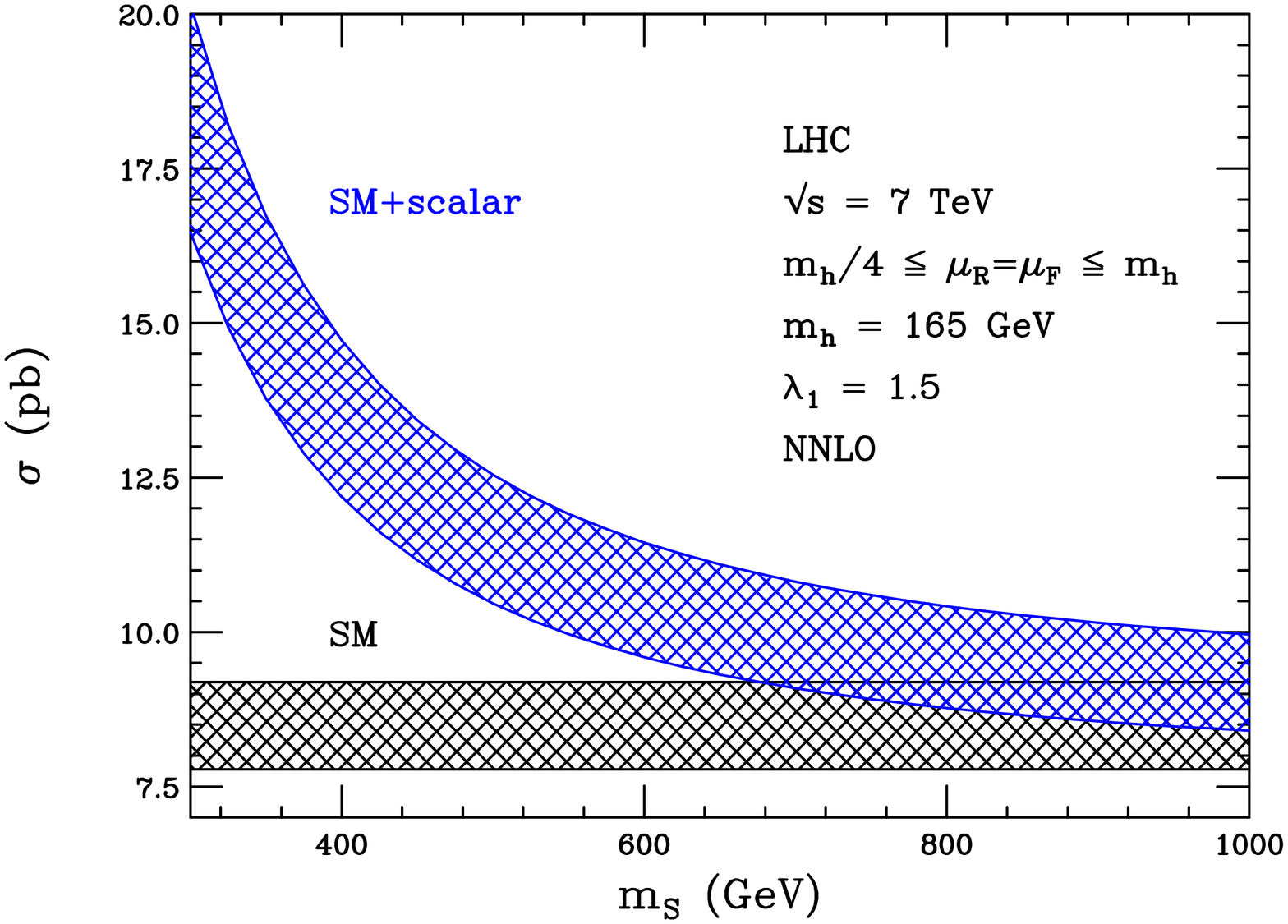}
   \vspace{-0.5cm}
   \caption{Higgs production cross section in gluon-fusion at the LHC for $m_h=165$ GeV as a function of the scalar mass.  The bands indicate the scale variation $m_h/4 \leq \mu \leq m_h$.  All other parameters are as described in the text.  Also shown is the SM cross section with its corresponding scale uncertainty.}
   \label{fig:LHCmsplot}
\end{figure}

It is interesting to speculate on the potential of the Tevatron experiments to exclude color-octet scalar parameter space using the established exclusion limit on Higgs boson production.  As the Tevatron limits include contributions from Higgs production in association with gauge bosons, and also from weak-boson fusion production of the Higgs, it is not possible to simply scale current exclusion limits by the ratio of the gluon-fusion production cross section with the scalar over that of the SM.  The Tevatron collaborations plan to release limits utilizing only the gluon-fusion mode to make this scaling possible~\cite{herndon}.  However, it is likely that scalar masses in excess of the expected direct-search limits can be probed.  As discussed in the previous section, the scalar contributions depend primarily on the ratio $\lambda_1 v^2 /(m_S^2)$.  It would be interesting to derive the allowed region in the plane of this ratio versus the Higgs boson mass.  One caveat is that the Higgs production cross section depends on $\lambda_1 / m_S^2$, while the pair production mode used in the direct search only depends on $m_S$.  The indirect constraint discussed here is therefore more model dependent.  Nevertheless, since there is no symmetry argument which suggests that $\lambda_1$ is small, the Higgs exclusion limit probes a large and relevant region of scalar parameter space.

To facilitate studies of color-octet scalar effects on the Higgs boson production cross section, we display in Table~\ref{tab:TEV1} and Table~\ref{tab:LHC1} the cross sections for Higgs production
for the example points $m_S = 300$ GeV and $m_S = 800$ GeV at the Tevatron and the LHC.  We include in this table the errors arising from both scale variation, and also from imprecise knowledge of parton distribution functions (PDFs).  For the PDF errors, we show the 90\% confidence-level error estimates derived using the MSTW error eigenvectors.  Other possible choices for the PDF uncertainty estimate could be used: either the 68\% confidence-level error, or the combined PDF+$\alpha_s$ error~\cite{Martin:2009bu}.  In each table we also show the fractional deviation of the cross section from the SM prediction, which we denote as $\delta$.  The deviations are roughly a factor of two for $m_S=300$ GeV and $20\%$ for $m_S=800$ GeV.  At the Tevatron both the scale and PDF errors are roughly $\pm10\%$.  At the LHC the PDF error reduces to a few percent.  For both mass points the scalar-induced deviations from the SM are larger than the combined errors.

\begin{table}[htbp]
  \begin{center}
    \begin{tabular}{|c|c|c||c|c|c|}
    \hline
    \multicolumn{3}{|c||}{$m_S = 300$ GeV} &  \multicolumn{3}{c|}{$m_S = 800$ GeV} \\
      \hline \hline
      		$m_h$ (GeV) & $\sigma$ (pb) & $\delta$ & $m_h$ (GeV) & $\sigma$ (pb) & $\delta$ \\ \hline

115 & $ 3.901^{+0.483(sc)+0.234(pdf)}_{-0.542(sc)-0.254(pdf)}$ &  2.191 & 115 & $ 1.501^{+0.131(sc)+0.090(pdf)}_{-0.182(sc)-0.098(pdf)}$ &  0.228 \\ \hline
120 & $ 3.418^{+0.421(sc)+0.213(pdf)}_{-0.475(sc)-0.230(pdf)}$ &  2.187 & 120 & $ 1.316^{+0.115(sc)+0.082(pdf)}_{-0.159(sc)-0.089(pdf)}$ &  0.227 \\ \hline
125 & $ 3.006^{+0.368(sc)+0.195(pdf)}_{-0.418(sc)-0.209(pdf)}$ &  2.183 & 125 & $ 1.159^{+0.100(sc)+0.075(pdf)}_{-0.140(sc)-0.081(pdf)}$ &  0.227 \\ \hline
130 & $ 2.653^{+0.323(sc)+0.177(pdf)}_{-0.369(sc)-0.192(pdf)}$ &  2.179 & 130 & $ 1.023^{+0.088(sc)+0.068(pdf)}_{-0.124(sc)-0.074(pdf)}$ &  0.226 \\ \hline
135 & $ 2.348^{+0.285(sc)+0.162(pdf)}_{-0.327(sc)-0.175(pdf)}$ &  2.175 & 135 & $ 0.907^{+0.078(sc)+0.063(pdf)}_{-0.110(sc)-0.068(pdf)}$ &  0.226 \\ \hline
140 & $ 2.085^{+0.252(sc)+0.149(pdf)}_{-0.290(sc)-0.160(pdf)}$ &  2.170 & 140 & $ 0.806^{+0.069(sc)+0.057(pdf)}_{-0.098(sc)-0.062(pdf)}$ &  0.225 \\ \hline
145 & $ 1.857^{+0.223(sc)+0.136(pdf)}_{-0.259(sc)-0.147(pdf)}$ &  2.166 & 145 & $ 0.718^{+0.061(sc)+0.053(pdf)}_{-0.087(sc)-0.057(pdf)}$ &  0.224 \\ \hline
150 & $ 1.658^{+0.199(sc)+0.125(pdf)}_{-0.231(sc)-0.135(pdf)}$ &  2.162 & 150 & $ 0.641^{+0.054(sc)+0.049(pdf)}_{-0.078(sc)-0.052(pdf)}$ &  0.223 \\ \hline
155 & $ 1.484^{+0.177(sc)+0.115(pdf)}_{-0.207(sc)-0.125(pdf)}$ &  2.158 & 155 & $ 0.574^{+0.048(sc)+0.045(pdf)}_{-0.070(sc)-0.048(pdf)}$ &  0.223 \\ \hline
160 & $ 1.331^{+0.159(sc)+0.107(pdf)}_{-0.186(sc)-0.115(pdf)}$ &  2.154 & 160 & $ 0.516^{+0.043(sc)+0.041(pdf)}_{-0.063(sc)-0.044(pdf)}$ &  0.222 \\ \hline
165 & $ 1.197^{+0.142(sc)+0.099(pdf)}_{-0.167(sc)-0.106(pdf)}$ &  2.149 & 165 & $ 0.464^{+0.039(sc)+0.038(pdf)}_{-0.056(sc)-0.041(pdf)}$ &  0.221 \\ \hline
170 & $ 1.078^{+0.128(sc)+0.091(pdf)}_{-0.151(sc)-0.098(pdf)}$ &  2.145 & 170 & $ 0.418^{+0.035(sc)+0.035(pdf)}_{-0.051(sc)-0.038(pdf)}$ &  0.220 \\ \hline
175 & $ 0.974^{+0.115(sc)+0.084(pdf)}_{-0.136(sc)-0.091(pdf)}$ &  2.140 & 175 & $ 0.378^{+0.032(sc)+0.033(pdf)}_{-0.046(sc)-0.035(pdf)}$ &  0.220 \\ \hline
180 & $ 0.881^{+0.104(sc)+0.078(pdf)}_{-0.123(sc)-0.084(pdf)}$ &  2.136 & 180 & $ 0.342^{+0.029(sc)+0.030(pdf)}_{-0.042(sc)-0.033(pdf)}$ &  0.219 \\ \hline
185 & $ 0.799^{+0.094(sc)+0.072(pdf)}_{-0.112(sc)-0.078(pdf)}$ &  2.131 & 185 & $ 0.311^{+0.026(sc)+0.028(pdf)}_{-0.038(sc)-0.030(pdf)}$ &  0.218 \\ \hline
190 & $ 0.725^{+0.085(sc)+0.068(pdf)}_{-0.101(sc)-0.072(pdf)}$ &  2.126 & 190 & $ 0.282^{+0.023(sc)+0.026(pdf)}_{-0.035(sc)-0.028(pdf)}$ &  0.217 \\ \hline
195 & $ 0.660^{+0.078(sc)+0.063(pdf)}_{-0.092(sc)-0.067(pdf)}$ &  2.121 & 195 & $ 0.257^{+0.021(sc)+0.024(pdf)}_{-0.031(sc)-0.026(pdf)}$ &  0.216 \\ \hline
200 & $ 0.601^{+0.070(sc)+0.059(pdf)}_{-0.084(sc)-0.063(pdf)}$ &  2.116 & 200 & $ 0.234^{+0.019(sc)+0.023(pdf)}_{-0.029(sc)-0.024(pdf)}$ &  0.215 \\ \hline

		 \hline
          \end{tabular}
  \end{center}
  \caption{      Table of cross sections for $m_S=300$ and $800$ GeV at the Tevatron.  Values are shown for Higgs masses between 115 and 200 GeV, and for $\lambda_1=2.5$..  The scale and PDF errors are denoted by the labels $sc$ and $pdf$, respectively.  The fractional deviations of each cross section from the SM for the scale choice $\mu=m_h/2$ are indicated by $\delta$. \label{tab:TEV1} }
\end{table}

\begin{table}[htbp]
  \begin{center}
    \begin{tabular}{|c|c|c||c|c|c|}
    \hline
    \multicolumn{3}{|c||}{$m_S = 300$ GeV} &  \multicolumn{3}{c|}{$m_S = 800$ GeV} \\
      \hline \hline
      		$m_h$ (GeV) & $\sigma$ (pb) & $\delta$ & $m_h$ (GeV) & $\sigma$ (pb) & $\delta$ \\ \hline

115 & $ 40.08^{+ 4.24(sc)+ 1.05(pdf)}_{- 4.23(sc)- 1.28(pdf)}$ &   1.17 & 115 & $ 20.99^{+ 1.80(sc)+ 0.55(pdf)}_{- 1.99(sc)- 0.67(pdf)}$ &  0.134 \\ \hline
125 & $ 33.60^{+ 3.47(sc)+ 0.90(pdf)}_{- 3.53(sc)- 1.09(pdf)}$ &   1.16 & 125 & $ 17.61^{+ 1.47(sc)+ 0.47(pdf)}_{- 1.66(sc)- 0.57(pdf)}$ &  0.133 \\ \hline
135 & $ 28.49^{+ 2.89(sc)+ 0.77(pdf)}_{- 2.99(sc)- 0.95(pdf)}$ &   1.16 & 135 & $ 14.95^{+ 1.22(sc)+ 0.41(pdf)}_{- 1.40(sc)- 0.49(pdf)}$ &  0.132 \\ \hline
145 & $ 24.40^{+ 2.43(sc)+ 0.68(pdf)}_{- 2.55(sc)- 0.83(pdf)}$ &   1.15 & 145 & $ 12.82^{+ 1.02(sc)+ 0.36(pdf)}_{- 1.20(sc)- 0.43(pdf)}$ &  0.132 \\ \hline
155 & $ 21.09^{+ 2.07(sc)+ 0.61(pdf)}_{- 2.20(sc)- 0.73(pdf)}$ &   1.15 & 155 & $ 11.09^{+ 0.87(sc)+ 0.32(pdf)}_{- 1.03(sc)- 0.38(pdf)}$ &  0.131 \\ \hline
165 & $ 18.38^{+ 1.78(sc)+ 0.55(pdf)}_{- 1.91(sc)- 0.65(pdf)}$ &   1.15 & 165 & $  9.67^{+ 0.75(sc)+ 0.29(pdf)}_{- 0.90(sc)- 0.34(pdf)}$ &  0.130 \\ \hline
175 & $ 16.13^{+ 1.55(sc)+ 0.50(pdf)}_{- 1.68(sc)- 0.58(pdf)}$ &   1.14 & 175 & $  8.50^{+ 0.65(sc)+ 0.26(pdf)}_{- 0.79(sc)- 0.31(pdf)}$ &  0.129 \\ \hline
185 & $ 14.26^{+ 1.35(sc)+ 0.46(pdf)}_{- 1.49(sc)- 0.54(pdf)}$ &   1.14 & 185 & $  7.52^{+ 0.57(sc)+ 0.24(pdf)}_{- 0.70(sc)- 0.28(pdf)}$ &  0.128 \\ \hline
195 & $ 12.69^{+ 1.19(sc)+ 0.42(pdf)}_{- 1.32(sc)- 0.49(pdf)}$ &   1.14 & 195 & $  6.70^{+ 0.50(sc)+ 0.23(pdf)}_{- 0.62(sc)- 0.25(pdf)}$ &  0.127 \\ \hline
205 & $ 11.35^{+ 1.05(sc)+ 0.39(pdf)}_{- 1.18(sc)- 0.45(pdf)}$ &   1.13 & 205 & $  6.00^{+ 0.44(sc)+ 0.21(pdf)}_{- 0.56(sc)- 0.24(pdf)}$ &  0.127 \\ \hline
215 & $ 10.22^{+ 0.94(sc)+ 0.37(pdf)}_{- 1.06(sc)- 0.41(pdf)}$ &   1.12 & 215 & $  5.41^{+ 0.39(sc)+ 0.19(pdf)}_{- 0.50(sc)- 0.22(pdf)}$ &  0.125 \\ \hline
225 & $  9.25^{+ 0.84(sc)+ 0.34(pdf)}_{- 0.96(sc)- 0.38(pdf)}$ &   1.12 & 225 & $  4.90^{+ 0.35(sc)+ 0.18(pdf)}_{- 0.45(sc)- 0.20(pdf)}$ &  0.124 \\ \hline
235 & $  8.41^{+ 0.76(sc)+ 0.32(pdf)}_{- 0.87(sc)- 0.36(pdf)}$ &   1.11 & 235 & $  4.47^{+ 0.32(sc)+ 0.17(pdf)}_{- 0.41(sc)- 0.19(pdf)}$ &  0.123 \\ \hline
245 & $  7.69^{+ 0.69(sc)+ 0.30(pdf)}_{- 0.80(sc)- 0.34(pdf)}$ &   1.11 & 245 & $  4.09^{+ 0.29(sc)+ 0.16(pdf)}_{- 0.38(sc)- 0.18(pdf)}$ &  0.122 \\ \hline
255 & $  7.07^{+ 0.63(sc)+ 0.29(pdf)}_{- 0.73(sc)- 0.32(pdf)}$ &   1.10 & 255 & $  3.77^{+ 0.26(sc)+ 0.15(pdf)}_{- 0.35(sc)- 0.17(pdf)}$ &  0.120 \\ \hline
265 & $  6.53^{+ 0.58(sc)+ 0.28(pdf)}_{- 0.68(sc)- 0.30(pdf)}$ &   1.09 & 265 & $  3.49^{+ 0.24(sc)+ 0.15(pdf)}_{- 0.32(sc)- 0.16(pdf)}$ &  0.119 \\ \hline
275 & $  6.06^{+ 0.54(sc)+ 0.26(pdf)}_{- 0.63(sc)- 0.29(pdf)}$ &   1.08 & 275 & $  3.25^{+ 0.23(sc)+ 0.14(pdf)}_{- 0.30(sc)- 0.15(pdf)}$ &  0.117 \\ \hline
285 & $  5.66^{+ 0.50(sc)+ 0.25(pdf)}_{- 0.59(sc)- 0.28(pdf)}$ &   1.07 & 285 & $  3.05^{+ 0.21(sc)+ 0.14(pdf)}_{- 0.28(sc)- 0.15(pdf)}$ &  0.115 \\ \hline
295 & $  5.31^{+ 0.47(sc)+ 0.24(pdf)}_{- 0.55(sc)- 0.27(pdf)}$ &   1.06 & 295 & $  2.87^{+ 0.20(sc)+ 0.13(pdf)}_{- 0.27(sc)- 0.14(pdf)}$ &  0.114 \\ \hline
305 & $  5.02^{+ 0.44(sc)+ 0.24(pdf)}_{- 0.52(sc)- 0.26(pdf)}$ &   1.05 & 305 & $  2.72^{+ 0.19(sc)+ 0.13(pdf)}_{- 0.25(sc)- 0.14(pdf)}$ &  0.111 \\ \hline
315 & $  4.77^{+ 0.42(sc)+ 0.23(pdf)}_{- 0.49(sc)- 0.25(pdf)}$ &   1.03 & 315 & $  2.61^{+ 0.18(sc)+ 0.13(pdf)}_{- 0.24(sc)- 0.14(pdf)}$ &  0.109 \\ \hline
325 & $  4.58^{+ 0.40(sc)+ 0.23(pdf)}_{- 0.47(sc)- 0.25(pdf)}$ &   1.01 & 325 & $  2.52^{+ 0.17(sc)+ 0.13(pdf)}_{- 0.23(sc)- 0.14(pdf)}$ &  0.107 \\ \hline
335 & $  4.45^{+ 0.38(sc)+ 0.23(pdf)}_{- 0.46(sc)- 0.25(pdf)}$ &   0.99 & 335 & $  2.47^{+ 0.17(sc)+ 0.13(pdf)}_{- 0.23(sc)- 0.14(pdf)}$ &  0.104 \\ \hline
345 & $  4.43^{+ 0.38(sc)+ 0.23(pdf)}_{- 0.46(sc)- 0.25(pdf)}$ &   0.95 & 345 & $  2.50^{+ 0.17(sc)+ 0.13(pdf)}_{- 0.23(sc)- 0.14(pdf)}$ &  0.100 \\ \hline
355 & $  4.55^{+ 0.39(sc)+ 0.24(pdf)}_{- 0.47(sc)- 0.26(pdf)}$ &   0.90 & 355 & $  2.62^{+ 0.18(sc)+ 0.14(pdf)}_{- 0.24(sc)- 0.15(pdf)}$ &  0.094 \\ \hline
365 & $  4.52^{+ 0.39(sc)+ 0.25(pdf)}_{- 0.47(sc)- 0.27(pdf)}$ &   0.87 & 365 & $  2.64^{+ 0.18(sc)+ 0.14(pdf)}_{- 0.24(sc)- 0.16(pdf)}$ &  0.091 \\ \hline
375 & $  4.39^{+ 0.37(sc)+ 0.24(pdf)}_{- 0.45(sc)- 0.27(pdf)}$ &   0.85 & 375 & $  2.58^{+ 0.17(sc)+ 0.14(pdf)}_{- 0.24(sc)- 0.16(pdf)}$ &  0.088 \\ \hline
385 & $  4.19^{+ 0.36(sc)+ 0.24(pdf)}_{- 0.43(sc)- 0.26(pdf)}$ &   0.84 & 385 & $  2.47^{+ 0.16(sc)+ 0.14(pdf)}_{- 0.23(sc)- 0.15(pdf)}$ &  0.086 \\ \hline
395 & $  3.96^{+ 0.34(sc)+ 0.23(pdf)}_{- 0.41(sc)- 0.25(pdf)}$ &   0.83 & 395 & $  2.34^{+ 0.15(sc)+ 0.14(pdf)}_{- 0.22(sc)- 0.15(pdf)}$ &  0.085 \\ \hline
405 & $  3.71^{+ 0.31(sc)+ 0.22(pdf)}_{- 0.38(sc)- 0.24(pdf)}$ &   0.83 & 405 & $  2.19^{+ 0.14(sc)+ 0.13(pdf)}_{- 0.20(sc)- 0.14(pdf)}$ &  0.084 \\ \hline
415 & $  3.45^{+ 0.29(sc)+ 0.21(pdf)}_{- 0.36(sc)- 0.23(pdf)}$ &   0.84 & 415 & $  2.04^{+ 0.13(sc)+ 0.12(pdf)}_{- 0.19(sc)- 0.13(pdf)}$ &  0.082 \\ \hline

		 \hline
          \end{tabular}
  \end{center}
  \caption{      Table of cross sections for $m_S=300$ and $800$ GeV at the LHC.  Values are shown for Higgs masses between 115 and 200 GeV, and for $\lambda_1=1.5$.  The scale and PDF errors are denoted by the labels $sc$ and $pdf$, respectively.  The fractional deviations of each cross section from the SM for the scale choice $\mu=m_h/2$ are indicated by $\delta$. \label{tab:LHC1} }
\end{table}

\section{Conclusions}
\label{sec:conc}

In this paper we have derived the NNLO Wilson coefficient which describes the effect of an ${\bf (8,1)_0}$ scalar on the Higgs-gluon effective Lagrangian.  We have presented simple analytic formulae for the Wilson coefficient and all required renormalization constants for use in future studies.  Our calculation revealed an interesting technical feature of scalar contributions to the $gg \to h$ cross section which first appears at NNLO.  At this order, the quartic-scalar potential must be included in order to properly renormalize the Wilson coefficient.  In our example with a single ${\bf (8,1)_0}$ scalar, only one additional Lagrangian term was required.  In theories with a more complicated scalar potential, such as that for a ${\bf (8,2)_{1/2}}$ scalar~\cite{Manohar:2006ga}, the addition of multiple operators is in principle required to obtain a finite NNLO result.

We have performed a renormalization-group analysis of the scalar-sector couplings to determine the theoretically consistent region in which perturbation theory can be applied.  We have also studied the phenomenological predictions for both the Tevatron and LHC.  The scalar-induced effects are large at both colliders, and are visible over both scale and PDF errors for scalar masses approaching 1 TeV.  This suggests that the scalar parameters can be stringently constrained using the exclusion limit 
on the Higgs boson production cross section obtained by the Tevatron, and that the limits would be competitive with those obtained from direct searches for scalar pair production.  If the gluon-fusion contribution to the Tevatron exclusion limit were separately presented, it would be possible to rescale these results to obtain bounds 
on the $\lambda_1 /m_S^2$ combination of scalar parameters.

In summary, color-octet scalars are an interesting and phenomenologically-rich example of physics beyond the SM.  The NNLO calculation presented here quantifies precisely the effect of a ${\bf (8,1)_{0}}$ scalar on the Higgs production cross section in gluon fusion, and our phenomenological study shows that the deviations induced by the scalar are large at the Tevatron and the LHC.  The exclusion limit on Higgs production set by the Tevatron collaborations allows for the scalar parameters 
to be stringently constrained, and we hope that this analysis is undertaken.

\section*{Acknowledgments}

R.~B. is supported by the Swiss National Science Foundation under contract 200020-116756/2.  F.P. is supported by the U.S. Department of Energy, Division of High Energy Physics, under contract DE-AC02-06CH11357 and the grant DE-FG02-95ER40896.


\begin{thebibliography}{99}

\bibitem{Aaltonen:2010yv}
  T.~Aaltonen {\it et al.}  [CDF and D0 Collaborations],
  Phys.\ Rev.\ Lett.\  {\bf 104}, 061802 (2010)
  [arXiv:1001.4162 [hep-ex]].

\bibitem{Barate:2003sz}
  R.~Barate {\it et al.}  [LEP Working Group for Higgs boson searches and
                  ALEPH Collaboration and  and],
  Phys.\ Lett.\  B {\bf 565}, 61 (2003)
  [arXiv:hep-ex/0306033].

\bibitem{Djouadi:1991tka}
  A.~Djouadi, M.~Spira and P.~M.~Zerwas,
  Phys.\ Lett.\  B {\bf 264}, 440 (1991).
  
  \bibitem{Spira:1995rr}
  M.~Spira, A.~Djouadi, D.~Graudenz and P.~M.~Zerwas,
  Nucl.\ Phys.\  B {\bf 453}, 17 (1995)
  [arXiv:hep-ph/9504378].

\bibitem{Dawson:1990zj}
  S.~Dawson,
  Nucl.\ Phys.\  B {\bf 359}, 283 (1991).

\bibitem{Harlander:2002wh}
  R.~V.~Harlander and W.~B.~Kilgore,
  Phys.\ Rev.\ Lett.\  {\bf 88}, 201801 (2002)
  [arXiv:hep-ph/0201206].

\bibitem{Anastasiou:2002yz}
  C.~Anastasiou and K.~Melnikov,
  Nucl.\ Phys.\  B {\bf 646}, 220 (2002)
  [arXiv:hep-ph/0207004].
  
  \bibitem{Ravindran:2003um}
  V.~Ravindran, J.~Smith and W.~L.~van Neerven,
  Nucl.\ Phys.\  B {\bf 665}, 325 (2003)
  [arXiv:hep-ph/0302135].

\bibitem{ztalks}
See the talks by G. Degrassi and R. Harlander at the Workshop on Higgs Boson Phenomenology, 7-9 January 2009, Zurich, Switzerland, at the 
site \verb+http://www.itp.uzh.ch/events/higgsboson2009/index.html+.  See also the recent papers 
R.~V.~Harlander and K.~J.~Ozeren,
  JHEP {\bf 0911}, 088 (2009)
  [arXiv:0909.3420 [hep-ph]];
 R.~V.~Harlander, H.~Mantler, S.~Marzani and K.~J.~Ozeren,
  arXiv:0912.2104 [hep-ph].

  \bibitem{Anastasiou:2008tj}
  C.~Anastasiou, R.~Boughezal and F.~Petriello,
  JHEP {\bf 0904}, 003 (2009)
  [arXiv:0811.3458 [hep-ph]].

\bibitem{deFlorian:2009hc}
  D.~de Florian and M.~Grazzini,
  Phys.\ Lett.\  B {\bf 674}, 291 (2009)
  [arXiv:0901.2427 [hep-ph]].

\bibitem{Boughezal:2009fw}
  R.~Boughezal,
  arXiv:0908.3641 [hep-ph].

\bibitem{Low:2009nj}
  For a recent analysis, see I.~Low and S.~Shalgar,
  JHEP {\bf 0904}, 091 (2009)
  [arXiv:0901.0266 [hep-ph]].

\bibitem{Hewett:2002nk}
  J.~L.~Hewett and T.~G.~Rizzo,
  JHEP {\bf 0308}, 028 (2003)
  [arXiv:hep-ph/0202155].

\bibitem{Manohar:2006ga}
  A.~V.~Manohar and M.~B.~Wise,
  Phys.\ Rev.\  D {\bf 74}, 035009 (2006)
  [arXiv:hep-ph/0606172].

\bibitem{Bonciani:2007ex}
  R.~Bonciani, G.~Degrassi and A.~Vicini,
  JHEP {\bf 0711}, 095 (2007)
  [arXiv:0709.4227 [hep-ph]].

\bibitem{Dobrescu:2007xf}
  B.~A.~Dobrescu, K.~Kong and R.~Mahbubani,
  JHEP {\bf 0707}, 006 (2007)
  [arXiv:hep-ph/0703231].
  
  \bibitem{Dobrescu:2007yp}
  B.~A.~Dobrescu, K.~Kong and R.~Mahbubani,
  Phys.\ Lett.\  B {\bf 670}, 119 (2008)
  [arXiv:0709.2378 [hep-ph]].

\bibitem{Hill:2002ap}
  C.~T.~Hill and E.~H.~Simmons,
  Phys.\ Rept.\  {\bf 381}, 235 (2003)
  [Erratum-ibid.\  {\bf 390}, 553 (2004)]
  [arXiv:hep-ph/0203079].

\bibitem{Callaway:1988ya}
  D.~J.~E.~Callaway,
  Phys.\ Rept.\  {\bf 167}, 241 (1988).

\bibitem{Steinhauser:2002rq}
  M.~Steinhauser,
  Phys.\ Rept.\  {\bf 364} (2002) 247
  [arXiv:hep-ph/0201075].

\bibitem{Chetyrkin:1997un}
  K.~G.~Chetyrkin, B.~A.~Kniehl and M.~Steinhauser,
  Nucl.\ Phys.\  B {\bf 510}, 61 (1998)
  [arXiv:hep-ph/9708255].

\bibitem{Chetyrkin:1996ke}
  K.~G.~Chetyrkin, B.~A.~Kniehl and M.~Steinhauser,
  Nucl.\ Phys.\  B {\bf 490} (1997) 19
  [arXiv:hep-ph/9701277].

\bibitem{Spiridonov:1984br}
  V.~P.~Spiridonov,
  ``Anomalous Dimension of $G^2_{\mu\nu}$ and $\beta$-Function,''
  Report No. INR P-0378, Moscow, 1984.

\bibitem{Spiridonov:1988md}
  V.~P.~Spiridonov and K.~G.~Chetyrkin,
  Sov.\ J.\ Nucl.\ Phys.\  {\bf 47} (1988) 522
  [Yad.\ Fiz.\  {\bf 47} (1988) 818].

\bibitem{Chetyrkin:1997iv}
 K.~G.~Chetyrkin, B.~A.~Kniehl and M.~Steinhauser,
 Phys.\ Rev.\ Lett.\  {\bf 79} (1997) 353
 [arXiv:hep-ph/9705240].

\bibitem{Kramer:1996iq}
  M.~Kramer, E.~Laenen and M.~Spira,
  Nucl.\ Phys.\  B {\bf 511}, 523 (1998)
  [arXiv:hep-ph/9611272].

\bibitem{:2009ec}
    [Tevatron Electroweak Working Group and CDF Collaboration and D0 Collab],
  arXiv:0903.2503 [hep-ex].

\bibitem{Kuhn:2007vp}
  J.~H.~Kuhn, M.~Steinhauser and C.~Sturm,
  Nucl.\ Phys.\  B {\bf 778}, 192 (2007)
  [arXiv:hep-ph/0702103].

\bibitem{Anastasiou:2009kn}
  C.~Anastasiou, S.~Bucherer and Z.~Kunszt,
  JHEP {\bf 0910}, 068 (2009)
  [arXiv:0907.2362 [hep-ph]].

\bibitem{Aglietti:2004nj}
  U.~Aglietti, R.~Bonciani, G.~Degrassi and A.~Vicini,
  Phys.\ Lett.\  B {\bf 595}, 432 (2004)
  [arXiv:hep-ph/0404071].

\bibitem{Actis:2008ug}
  S.~Actis, G.~Passarino, C.~Sturm and S.~Uccirati,
  Phys.\ Lett.\  B {\bf 670}, 12 (2008)
  [arXiv:0809.1301 [hep-ph]].
    
  \bibitem{Keung:2009bs}
  W.~Y.~Keung and F.~J.~Petriello,
  Phys.\ Rev.\  D {\bf 80}, 013007 (2009)
  [arXiv:0905.2775 [hep-ph]].

\bibitem{Martin:2009iq}
  A.~D.~Martin, W.~J.~Stirling, R.~S.~Thorne and G.~Watt,
  Eur.\ Phys.\ J.\  C {\bf 63}, 189 (2009)
  [arXiv:0901.0002 [hep-ph]].

\bibitem{Kniehl:1995tn}
  For a review, see B.~A.~Kniehl and M.~Spira,
  Z.\ Phys.\  C {\bf 69}, 77 (1995)
  [arXiv:hep-ph/9505225].

\bibitem{Bekavac:2009gz}
  S.~Bekavac, A.~G.~Grozin, D.~Seidel and V.~A.~Smirnov,
  Nucl.\ Phys.\  B {\bf 819} (2009) 183
  [arXiv:0903.4760 [hep-ph]].

\bibitem{herndon}
M. Herndon, private communication.

\bibitem{Martin:2009bu}
  A.~D.~Martin, W.~J.~Stirling, R.~S.~Thorne and G.~Watt,
  Eur.\ Phys.\ J.\  C {\bf 64}, 653 (2009)
  [arXiv:0905.3531 [hep-ph]].

\end{thebibliography}
\end{document}